\documentclass[twocolumn,aps,superscriptaddress]{revtex4}
\usepackage{palatino}
\usepackage[latin1]{inputenc}
\usepackage{epsf}
\usepackage{amsmath,amssymb}
\usepackage{latexsym}
\usepackage{calc}
\usepackage{color}
\usepackage{shadow}
\usepackage{epsfig}

\def\bea{\begin{eqnarray}}
\def\eea{\end{eqnarray}}
\def\ben{\begin{equation}}
\def\een{\end{equation}}
\def\benu{\begin{enumerate}}
\def\enu{\end{enumerate}}

\def\n{n}

\def\sss{\scriptscriptstyle\rm}

\def\1var{(\bx_1...\bx\N)}

\def\br{{\bf r}}

\def\bx{{\br t}}

\def\x{_{\sss X}}
\def\c{_{\sss C}}
\def\s{_{\sss S}}
\def\xc{_{\sss XC}}

\def\N{_{\sss N}}
\def\H{_{\sss H}}

\def\ext{_{\rm ext}}

\begin{document}
\title{Universal Dynamical Steps in the Exact Time-Dependent Exchange-Correlation Potential}
\author{P. Elliott}
\address{Department of Physics and Astronomy, Hunter College and the Graduate Center of the City University of New York, 695 Park Avenue, New York, New York 10065, USA}
\author{J.I.~Fuks}
\address{Nano-Bio Spectroscopy group, Dpto.~F\'isica de
  Materiales, Universidad del Pa\'is Vasco, Centro de F\'isica de
  Materiales CSIC-UPV/EHU-MPC and DIPC, Av.~Tolosa 72, E-20018 San
  Sebasti\'an, Spain}
\author{A.~Rubio}
\address{Nano-Bio Spectroscopy group, Dpto.~F\'isica de
  Materiales, Universidad del Pa\'is Vasco, Centro de F\'isica de
  Materiales CSIC-UPV/EHU-MPC and DIPC, Av.~Tolosa 72, E-20018 San
  Sebasti\'an, Spain}
\address{Fritz-Haber-Institut der Max-Planck-Gesellschaft,
Faradayweg 4-6, D-14195 Berlin, Germany}
\author{N. T. Maitra}
\address{Department of Physics and Astronomy, Hunter College and the Graduate Center of the City University of New York, 695 Park Avenue, New York, New York 10065, USA}

\date{\today}

\begin{abstract}
We show that the exact exchange-correlation potential of
time-dependent density-functional theory displays
dynamical step structures that have a spatially non-local and
time non-local dependence on the density. Using one-dimensional two-electron
model systems, we illustrate these steps for a range
of non-equilibrium dynamical situations relevant for modeling of
photo-chemical/physical processes: field-free evolution of a
non-stationary state, resonant local excitation, resonant complete
charge-transfer, and evolution under an arbitrary field. Lack of these
steps in usual approximations yield inaccurate dynamics, for
example predicting faster dynamics and incomplete charge
transfer.

\end{abstract}
\maketitle 

The vast majority of applications of time-dependent density functional
theory (TDDFT) today deal with the calculation of the linear
electronic spectra and response of molecules and solids, and provide
an unprecedented balance between accuracy and
efficiency~\cite{RG84,TDDFTbook12}. The theorems of TDDFT also apply
to any real-time electron dynamics, not necessarily starting in a
ground-state, and possibly subject to strong or weak time-dependent
fields. Time-resolved dynamics are particularly important and topical
for TDDFT for two reasons. First, there is really no alternative
practical method for accurately describing correlated electron
dynamics, and second, many fascinating new phenomena and technological
applications lie in this realm. These include: attosecond control of
electron dynamics~\cite{KV08},  photo-induced coupled
electron-ion dynamics (for example in describing light-harvesting and
artificial photosyntheses), and 
photo-chemical/physical processes~\cite{TTRF08,GROT09} in general.  TDDFT in
theory yields all observables exactly, solely in terms of the
time-dependent density, however in practice, approximations must be
made both for the observable as a functional of the density, and for
the exchange-correlation (xc) functional. Thus the question arises as
to whether the approximate functionals that have been successful for
excitations predict equally well the dynamics in the more general
time-dependent context. In particular, the exact xc contribution to
the Kohn-Sham (KS) potential at time $t$  functionally
depends on the history of the density $n(\br,t'<t)$, the initial
interacting many-body state $\Psi_0$, and the choice of the initial KS
state $\Phi_0$: $v\xc[n;\Psi_0,\Phi_0](\br,t)$.  However, almost all
calculations today use an {\it adiabatic} approximation, $v\xc^A =
v\xc^{\rm g.s.}[n(t)]$, that inputs the instantaneous density into a
ground-state xc functional~\cite{HK64,PK03}, completely neglecting
both the history- and initial-state-dependence.  Further, the ground-state functional $v\xc^{\rm g.s.}$
must be approximated; hybrid functionals, that mix in a fraction of
exact-exchange to an xc functional that otherwise depends locally or
semi-locally in space on the density, are most popular for the spectra
of molecules, while the spatially local LDA and semi-local GGA's are
most popular for solids~(see Ref.~\cite{TDDFTbook12} and references
therein).

Although understanding when such approximations are expected to
work well or fail has advanced significantly in the linear response
regime~\cite{TDDFTbook12}, considerably less is known about the performance
of approximate TDDFT for general non-linear time-dependent dynamics~\cite{Baerissue,HFTA11, RN11}. Part of the
reason for this is due to the lack of exact, or highly accurate, results to compare with. 
Moreover, even in the case where an
accurate calculation is available, it is very complicated
to extract the exact xc potential (although see
Refs.~\cite{RG12,NRL12} for significant progress).  Thus, it is critical for 
the reliability of 
TDDFT for describing fundamental dynamical processes in the
applications mentioned earlier, to first test available xc
approximations on systems 
for which the exact xc potential can be extracted.
One  such case 
is that of two-electrons in a spin-singlet,
chosen to start in a KS single-Slater determinant. We
 show that, in this case, the usual adiabatic and semi-local
approximations {\it typically} fail to capture a critical and fundamental structure in
the exact correlation potential: a
time-dependent step, that has a spatially ultranonlocal and
non-adiabatic dependence on the density. This feature is missing in all available TDDFT approximations today.
Even the exact 
 adiabatic functional misses this dynamical  step structure.  This leads to
erroneous dynamics, e.g. faster time scales are observed in the
adiabatic approximations for examples where the step opposes the density evolution.

For two-electrons in a spin-singlet we choose, as is usually done, the
initial KS state as a doubly-occupied spatial orbital, $\phi(\br,t)$. 
Then the exact KS potential for a given density evolution can be
found easily~\cite{HMB02}. In one-dimension (1D),
we have
\ben
\label{vsexact}
v\s(x,t)= -\frac{(\partial_x n(x,t))^2}{8n^2(x,t)} +\frac{\partial^2_x n(x,t)}{4 n(x,t)} - \frac{ u^2(x,t)}{2} - \int^x\frac{\partial u(x',t)}{\partial t}dx'
\een 
where $u(x,t)=j(x,t)/n(x,t)$ is the local ``velocity'', $n(x,t)$ is the one-body density, and $j(x,t)$ is the current-density.
We numerically solve the exact
time-dependent Schr\"odinger equation for the two-electron interacting wavefunction,
 obtain $n(x,t)$ and $j(x,t)$, and insert them into Eq.~\ref{vsexact}. 
The exchange-potential in this case is simply minus half the Hartree potential, $v\x(x,t) =
-v\H(x,t)/2$, with $v\H(x,t) = \int w(x',x)n(x',t) dx'$, in terms of the two-particle interaction $w(x',x)$. Therefore,
we
can directly extract the correlation potential using
\ben
\label{vcdef}
v\c(x,t) = v\s(x,t) - v\ext(x,t) -v\H(x,t)/2\;,
\een
where $v\ext(x,t)$ is the external potential applied to the system.
The two electrons in all our 1D examples interact via the soft-Coulomb
interaction~\cite{JES88}, $w(x',x) =1/\sqrt{(x'-x)^2+1}$. We use atomic units throughout. 

We start the analysis with some
purely (or largely) two-state systems, in which the exact
interacting time-dependent wavefunction, $|\Psi(t)\rangle$, can be
expanded in a basis consisting of the ground-state,
$|\Psi_g(t)\rangle$, and the first excited singlet state,
$|\Psi_e(t)\rangle$: 
\ben
\label{2lvlPSI}
 |\Psi(t)\rangle = a_g(t) |\Psi_g\rangle  + a_e(t) |\Psi_e\rangle
\een
where $a_g(t)$ and $a_e(t)$ are coefficients given by:
\ben
i \partial_t  \left( {\begin{array}{c}
 a_g(t)\\
 a_e(t)  \\
 \end{array} } \right)=
\left( {\begin{array}{cc}
 E_g -d_{gg}{\cal E}(t)  &  -d_{eg}{\cal E}(t)  \\
-d_{eg}{\cal E}(t) & E_e -d_{ee}{\cal E}(t)  \\
 \end{array} } \right) \left( {\begin{array}{c}
 a_g(t)   \\
 a_e(t)
 \end{array} } \right)
\label{coeffsEOM}
\een
where $E_g$, $E_e$ are the energy eigenvalues of the two
states, $d_{ab} = \int \Psi_a^*(x_1,x_2)(x_1+x_2)\Psi_b(x_1,x_2)
dx_1dx_2 $ is the transition dipole moment and ${\cal E}(t) =
A\cos(\omega t) $ is an applied electric field of strength $A$ and
frequency $\omega$.  In the weak amplitude limit, with $\omega \gg |d_{eg}A| $ and $\omega$ close to the resonant frequency, this reduces to the textbook Rabi problem. When
on-resonance, the system oscillates from one state to the other over a Rabi cycle of period $T_R  = 2\pi/(|d_{eg}|A)$.
By solving Eq.~\ref{coeffsEOM} we can easily construct the current and density at any time, their time-derivatives,  and hence all pieces entering Eq. \ref{vsexact}.

\begin{figure}[h]
 \begin{center}
  \includegraphics[width=0.475\textwidth,clip]{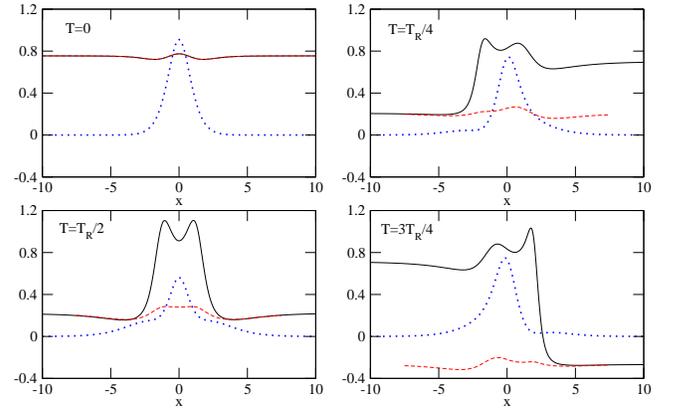} 
 \end{center}
 \caption{\label{f:vc-localex}(color online). Snapshots of the exact correlation potential (solid black),  density (blue dotted), and exact-adiabatic (red dashed) over one Rabi cycle; at $T_R/2$ the density of the first excited state is essentially exactly reached. In all graphs in this paper, the correlation potentials are plotted up to an arbitrary irrelevant time-dependent constant.}
\end{figure}
\begin{figure}[h]
 \begin{center}
  \includegraphics[width=0.5\textwidth,height=0.2\textwidth,clip]{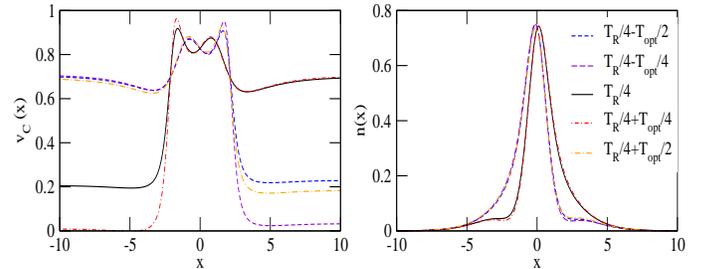} 
 \end{center}
 \caption{\label{f:vc-localexZoom}  (color online) Snapshots of the correlation potential (left), and corresponding density (right), at times indicated in the right panel. }
\end{figure}

In our first example, we consider a ``1D He atom'', where $v\ext =
-2/\sqrt{x^2 +1}$, subject to a weak electric field of strength $A=0.00667$au and
frequency $\omega = 0.533$au, resonant with the first singlet
excitation~\cite{RB09,FHTR11}. Figure~\ref{f:vc-localex} shows snapshots of the
correlation potential over one Rabi cycle, while
Fig.~\ref{f:vc-localexZoom} shows snapshots over one optical
period centered around $T_R/4$. (Note that the system is not exactly periodic over the Rabi period as the two time-scales dictated by the optical frequency and the Rabi frequency are not commensurate).

The most salient feature of the correlation potential is the presence of
time-dependent steps, that  oscillate on the
time-scale of the optical field. These steps arise from the fourth
term of Eq.~\ref{vsexact}: whenever there is a net ``acceleration'',
$\partial_tu(x,t)$, through the system, the spatial-integral is
finite, resulting in a potential rising from one end of the
system to the other. 
The correlation potential thus
has a spatially ultranonlocal dependence on the density, as it changes far from the system.

Further, the time-dependence of the steps is non-adiabatic, meaning that the instantaneous density is not enough to determine the
correlation potential functional. 
This is clear from  Figure 2 where the small changes of the density between time steps cannot capture the observed large changes of the step.
 One is tempted to point to the time-derivatives in the
fourth term in Eq.~\ref{vsexact} as further evidence for the non-adiabatic dependence, however
caution would be needed for such an argument as time non-locality in
$v\s$ is not the same as time non-locality in
$v\c$~\cite{TDDFTbook12}: the fourth term, for example, may be written
as $v\ext$ plus other terms, and although $v\ext$ has typically
strongly non-adiabatic dependence, this is irrelevant because it is
never approximated as a functional in practice~\cite{MLB08,TDDFTbook12}, rather it
is taken from the problem at hand. 
Only the xc potential must be approximated, and
the functional-dependence of this cannot be deduced directly from
Eq. (1).
Instead, to unambiguously show the
non-adiabatic dependence of the step, we plot the
``adiabatically-exact'' correlation potential in
Fig.~\ref{f:vc-localex}. This is defined by the exact
correlation potential for which both the interacting and KS wavefunctions
are ground-state wavefunctions with density equal to the instantaneous one
i.e. $v\c^{adia-ex}[\n] =
v\s^{adia}[\n]-v\ext^{adia}[\n]-v\H[n] -v\x[\n]$~\cite{TGK08}, where
$v\ext^{adia}[\n]$
is the external potential for two interacting
electrons whose ground-state has density $n$, and
$v\s^{adia}[\n]$ is the exact ground-state KS potential for this
density (given by the first two terms in Eq.(\ref{vsexact}). We find $v\ext^{adia}[\n]$ using similar techniques to
Ref. \onlinecite{TGK08} (see also Ref. \onlinecite{EM12} ). 
Fig.~\ref{f:vc-localex} shows that $v\c^{adia-ex}[\n]$ indeed
does  not capture the dynamical step structure.

Before turning to our next example, we verify that the two-state
approximation for the dynamics is accurate enough for our purposes.
Actually one aspect of the potentials we find is indeed an artifact of
the two-state approximation: the correlation potential asymptotically
has a slope so to exactly cancel the externally
applied electric field.  This is because the
two-state approximation cannot correctly describe polarization arising from occupying many excited states in time. The KS potential
obtained from inverting the two-state approximation must therefore be
flat asymptotically, as it cannot describe states that are polarized
asymptotically. The field is so weak in our case that this effect is
hardly noticeable on the scale of Figs.~\ref{f:vc-localex} and~\ref{f:vc-localexZoom}, but to check that our conclusions regarding the dynamics
step structure are unaffected by the two-level approximation, we computed
the KS potential using the density, current, and their time-derivatives
from the numerically exact wavefunction, found using
{\tt octopus}~\cite{octopus,octopus2}.  Apart from some extra structure in the tail region (small peaks and steps as we move away from the
atom), and the linear field-counteracting term, the correlation
potential agrees with that from the two-state
model.

Dynamical step features have arisen in TDDFT in earlier studies; 
Refs.~\cite{LK05,TGK08} showed they appear in ionization processes,
and linked them to a
time-dependent derivative discontinuity,
related to fractional charges. In time-resolved transport, step
structures have been shown to be essential for describing Coulomb-blockade
phenomena~\cite{KSKV10}, again related to the discontinuity.  In the response regime, field-counteracting
steps develop across long-range molecules~\cite{GSGB99}. In open-systems-TDDFT,
Ref.~\cite{TA11}  shows steps arise when using a closed KS system to model an
open interacting one. In the linear response regime, the xc kernel for charge-transfer excitations displays frequency-dependent steps~\cite{HG12}.
However we argue that the dynamical step structures we are seeing are
generic, and moreover, unlike most of the above cases~\cite{LK05,TGK08,KSKV10,GSGB99}, cannot be
captured by an adiabatic approximation. They appear with no need for
ionization nor subsystems of fractional charge, nor any 
applied field (see next example), unlike in
Refs.~\cite{LK05,TGK08,KSKV10,GSGB99}. In this sense our results are
more akin to Ref.~\cite{RG12}, which studies the physically very
different situation when an electron freely propagates through a
wire. The range of the examples we present suggests that such
non-adiabatic and non-local steps {\it generically} arise when dealing with real electron dynamics.

\begin{figure}[t]
 \begin{center}
  \includegraphics[width=0.45\textwidth,height=0.3\textwidth,clip]{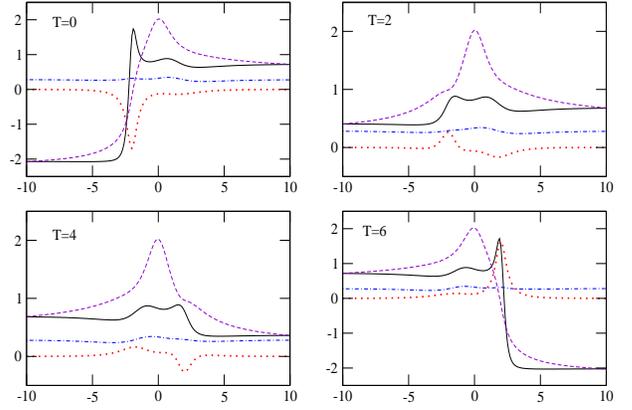} 
 \end{center}
 \caption{\label{f:IS_vc_adia} (color online) The exact correlation potential (black solid) shown at times 0, 2, 4, and 6 au, for the two-state example (Eq.~\ref{eq:IS}). Also shown are the local acceleration (red dotted), $\int^x \partial_t u - v\ext$ (purple dashed), and the adiabatically-exact correlation potential (blue dash-dot).}
\end{figure}

Our second example accentuates the fact that dynamical step structures
need neither ionization nor an external field to appear.  We begin in
an equal linear superposition of the ground and first-excited state of the 1D He and let it evolve freely, so that
\ben
|\Psi(t)\rangle = \left(e^{-iE_g t}|\Psi_g\rangle + e^{-iE_e t}|\Psi_e\rangle \right)/\sqrt{2}\;.
\label{eq:IS}
\een
It will oscillate back and forth between the two states with frequency
$\omega_0 = E_e - E_g$. The two-state approximation for this is exact at all times.
Again, we see large steps in the correlation potential, as
shown in Fig. \ref{f:IS_vc_adia}. To support the discussion
and provide a microscopic insight behind this phenomenon, we also plot in Fig. 3 
the acceleration, $a(x,t) = \partial_t u(x,t)$, and its spatial
integral with the external potential subtracted out.
The position and magnitude of the step at each time is heavily dependent on this
term. Peaks in the acceleration, when integrated, become local steps in
the potential and the asymptotic value of the step in $v\c(x,t)$ is
given by the total step in the spatial integral of
$a(x,t)$. Although local step-like features may be cancelled out by the other terms in
Eq. (\ref{vsexact}), the net magnitude of the step is
determined from the asymptotic values of this integral.

Note that we have the freedom to choose the initial state of the
KS system as long as it has the same density and first
derivative in time of the exact density~\cite{L99}, and the shape of the exact correlation potential depends on this choice~\cite{EM12}. We used a doubly occupied orbital
in the previous example, so that we can calculate the exact $v\c$ using the method
discussed. A different choice, with a configuration more
similar to that of the interacting initial state could well yield a
more gentle correlation potential~\cite{EM12}, with less dramatic step
structure.

\begin{figure}[t]
 \begin{center}
  \includegraphics[width=0.2\textwidth,height=0.45\textwidth,angle=270,clip]{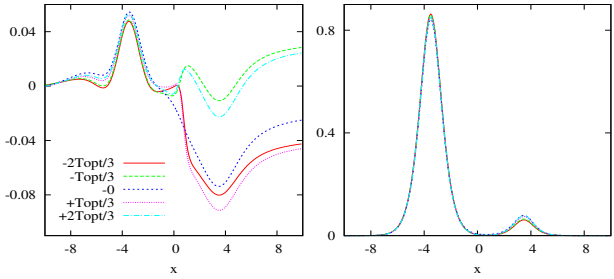} 
 \end{center}
 \caption{\label{f:CT_vc_strobe} (color online) The correlation potential (left) and density (right) shown at snapshots of time $T_R/8 \pm$ fractions of the optical period, $T_{\rm opt}  = 2\pi/0.112 $au, that are indicated in the key, for the two-well potential model Eq.~(\ref{eq:vext-ct}).}
\end{figure}

The generality of the step feature in dynamics is further supported by considering different
resonant excitations. Consider a double-well as a model of a
molecule:
\ben
v\ext(x,t) = -\frac{2}{\sqrt{(x+3.5)^2+1.0}} - \frac{1}{\cosh^2(x-3.5)} -{\cal E}(t)x
\label{eq:vext-ct}
\een
with ${\cal E}(t) = 0.006\cos(0.112 t)$. Here the ground-state has two electrons in one well, and a charge-transfer excited state $\Psi_e$, with one electron in each well,  at a frequency of 0.112au. We use the ground-state and  $\Psi_e$ 
 in the two-state model of Eq. (\ref{2lvlPSI}), and solve for the
occupations using Eq. (\ref{coeffsEOM}); we again checked the two-state result against the exact numerical solution using octopus. 
The system behaves like the Rabi problem with non-zero dipole moment for the ground-state~\cite{KM85,BMT00}.
In
Fig. \ref{f:CT_vc_strobe} we plot the correlation potential for several
times within an optical cycle around $T_R/8$. Again  dynamical steps oscillating on the
optical frequency time scale emerge. The situation is more complicated as a
step related to the delocalization of the density during the charge-transfer process slowly develops (on the time-scale of $T_R/2$)~\cite{FERM12}. The dynamical step can then increase,
decrease, or even reverse this charge-transfer step. 
Approximations unable to develop steps lead to incomplete charge-transfer.
This, 
along with other details of time resolved charge-transfer, is
investigated in more detail in Ref.~\cite{FERM12}. For
our purposes, it is sufficient to note that dynamical steps are again
present to capture the exact dynamics.

\begin{figure}[t]
 \begin{center}
  \includegraphics[width=0.5\textwidth,clip]{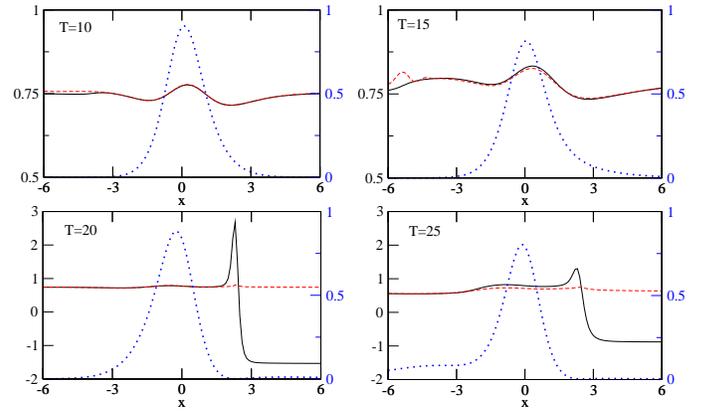} 
 \end{center}
 \caption{\label{f:ARB_vc}(color online). The exact correlation potentials (solid, black) at times $10$,$15$,$20$,$25$au during propagation under $\mathcal{E}(t) =
\frac{t}{2\sqrt{2}}\frac{0.3}{4\pi}\cos(0.3 \ t)$.  Also shown are the adiabatically-exact correlation potentials (dashed red), and the density (dotted blue, scale on the right).}
\end{figure}

Finally, we explicitly demonstrate that the non-local non-adiabatic
step feature is a generic aspect of the correlation potential in the following way.
 We subject the 1D
He atom to an electric field  that is chosen somewhat
arbitrarily: it is relatively strong and linearly switched on over 2 optical
cycles, with an off-resonant frequency.
In
Fig.~\ref{f:ARB_vc}, we show the exact correlation potential at four
times, along with the density, and the
adiabatically-exact correlation potential. The time-dependent step in
the exact $v\c$ is once again  evident, and again fails to be
captured by the adiabatically-exact approximation.

In summary, dynamical steps in the correlation potential are a generic feature of electron
dynamics. The step features discussed above arise from part of the
fourth term of Eq.~(\ref{vsexact}), which suggests that any time
there is a net localized acceleration, there will be a step, and that
it will have a very non-local spatial dependence on the density, and is
non-adiabatic. This represents a type of time-dependent screening, where the electron-electron interaction 
hinders electron movement to certain regions.
Although two-electron systems were studied here, we expect steps are a more general feature of electron dynamics, as supported by the recent Ref.~\cite{RG12}.

\begin{figure}[t]
 \begin{center}
  \includegraphics[width=0.4\textwidth,height=0.2\textwidth,clip]{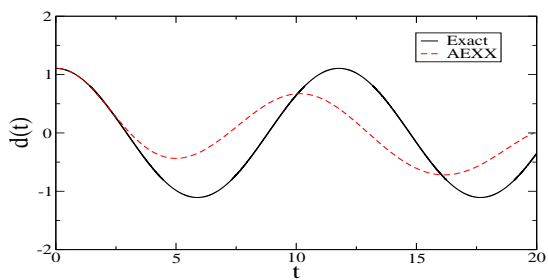} 
 \end{center}
 \caption{\label{f:IS_dip}(color online). The dipole moments in the field-free propagation of the linear-superposition state (see text),
where the TDKS calculation starts in the exact singlet doubly occupied orbital initial state.}
\end{figure}

The lack of the step in approximations leads to incorrect
dynamics. For example, faster time scales in adiabatic approximations
were found for the field-free dynamics of a linear superposition
state, where the direction of the step tended to oppose the density's
motion. The exact dipole and adiabatic exact-exchange (AEXX) dipole for this
case are shown in Fig.~\ref{f:IS_dip}.  We computed the dynamics of the
the local excitation in Figs 1 and 2 using AEXX, adiabatic LDA, and
adiabatic self-interaction-corrected LDA. In all cases, we found the
timescale for the dipole oscillations was faster than in the true
case, in spite of providing good linear response spectra~\cite{FHTR11}.  How general this finding is will be investigated in future work.

We note that the xc electric field, defined as the gradient
of the xc potential, has a more local character than the
potential. This suggests that considering functional approximations to
this field, or, more generally, to an xc vector
potential~\cite{VK96,RG12}, may point to an easier path to develop
approximations containing step features.

As applications of
TDDFT continue to expand, it is crucial to further study what the impact of the missing steps in
the approximations are on the predictions of these calculations.
When starting in the ground-state, the exact adiabatic potential may
follow well the exact dynamics at short times, but as soon as there is
an appreciable change in the occupation of an excited state, the exact
soution develops the dynamical step, entirely missing in the adiabatic
one. This result is general and applies to all available functionals. 
It raises an important issue when applying TDDFT to fundamental
photo-induced processes (e.g. photovoltaics, artificial
photosynthesis, photoactivated chemistry, photophysics, etc): all
these involve a significant change of state population. Clearly
population of many-body states due to the external field is not a
linear process and requires functionals able to cope with the generic
features of the dynamical step that we have unveiled in the present
work.



{\it Acknowledgments} Financial support from the National Science
Foundation (CHE-1152784), and a grant of computer time from the CUNY
High Performance Computing Center under NSF Grants CNS-0855217 and
CNS-0958379, are gratefully acknowledged. JIF acknowledges support
from an FPI-fellowship (FIS2007-65702-C02-01). AR acknowledge
financial support from the European Research Council Advanced Grant
DYNamo (ERC-2010-AdG -Proposal No. 267374) Spanish Grants
(FIS2011-65702- C02-01 and PIB2010US-00652), Grupos Consolidados
UPV/EHU del Gobierno Vasco (IT-319-07), and European Commission
project CRONOS (280879-2)

\end{document}